# Semantic network analysis of abstract and concrete word associations.

Dounia Lakhzoum, Marie Izaute, and Ludovic Ferrand

Université Clermont Auvergne, CNRS, LAPSCO, F-63001 Clermont-Ferrand, France

**Corresponding author**: Dounia Lakhzoum or Ludovic Ferrand, Université Clermont Auvergne, CNRS LAPSCO, 34 avenue Carnot TSA 60401- 63001 Clermont-Ferrand Cedex 1, France

Email: dounia.lakhzoum@ext.uca.fr

Author's note

This research was financed by the French government IDEX-ISITE initiative 16-IDEX-0001 (CAP 20-25).

ACKNOWLEGMENT

This work was sponsored by a public grant overseen by the French National Research Agency as part of the "Investissements d'Avenir" through the IDEX-ISITE initiative CAP 20-25 (ANR-16-IDEX-0001).

# ABSTRACT

In recent years, a new interest for the use of graph-theory based networks has emerged within the field of cognitive science. This has played a key role in mining the large amount of data generated by word association norms. In the present work, we applied semantic network analyses to explore norms of French word associations for concrete and abstract concepts (Lakhzoum et al., 2021). Graph analyses have shown that the network exhibits high clustering coefficient, sparse density, and small average shortest path length for both the concrete and abstract networks. These characteristics are consistent with a small-world structure. Comparisons between local node statistics and global structural topology showed that abstract and concrete concepts present a similar local connectivity but different overall patterns of structural organisation with concrete concepts presenting an organisation in densely connected communities compared to abstract concepts. These patterns confirm previously acquired knowledge about the dichotomy of abstract and concrete concepts on a larger scale. To the best of our knowledge, this is the first attempt to confirm the generalisability of these properties to the French language and with an emphasis on abstract and concrete concepts.

# INTRODUCTION

In recent years, a new interest has emerged within the field of cognitive science for mathematical models such as graph-theory based science networks (Baronchelli, Ferrer-i-Cancho, Pastor-Satorras, Chater, & Christiansen, 2013; see Siew, Wulff, Beckage, & Kenett, 2019 for a review). More specifically, the use of semantic network analysis on word association offers a powerful computational technique for the modelling of structures like the mental lexicon (De Deyne & Storms, 2008; De Deyne, Verheyen & Storms, 2016; De Deyne, Navarro, Perfors, Brysbaert & Storms, 2019; Stella, Beckage, Brede, & De Domenico, 2018). Models have shown that word associations are organised according to the same small-world structure found in many natural phenomena (Steyvers & Tenenbaum, 2005).

## Network science and characteristics

A network is built as a graph that consists of nodes linked by edges or arcs. In a network based on association data, the nodes correspond to words connected by an undirected edge or a directed arc. A directed network provides additional information about the direction of the association from cue to response words. The arcs and edges can have weights to represent the frequency of association. Adjacent nodes are directly connected by an arc or edge whereas nonadjacent nodes are connected by a path. Based on this macroscopic structure, it is possible to compute network statistics of interest. The macrostructure of the network can be described in terms of its diameter *(D)*, which gives the largest path length between two nodes. The average shortest path length *(ASPL)* that links two nodes is also of interest. It refers to the average minimum number of steps from one node to another. Nodes are characterised by the degree corresponding to their number of connections. In a directed network, each node has an in-degree $(k^{in})$ that is the number of incoming arcs and an out-degree $(k^{out})$ that is the number of outgoing arcs. An undirected network only has degree *k* corresponding to the number of edges of a node. The average degree *<k>* of the network can provide valuable information about the connectivity of the network itself but also about nodes centrality within the network. The clustering coefficient *(CC)* constitutes such a measure of centrality. It is the probability that two nodes are interconnected knowing they both are connected to a common neighbour. Finally, the density *(d)* is defined as a ratio of the number of edges in the network to the number of all possible edges (Steyvers & Tenenbaum, 2005; De Deyne & Storms, 2008; Csardi & Nepusz, 2006; Kolaczyk & Csardi, 2014; see Table 1 for a summary of all network statistics described above).

| network characteristics | Description |
| --- | --- |
| D | the diameter of the network |
| ASLP | the average length of shortest path |
| CC | the clustering coefficient |
| *<k>* | average degree |
| $k, k^{in}, k^{out}$ | the degree, in-degree, and out-degree |
| d | density of the network |
| S | Smallworldness |

**Table 1.** Description of network characteristics used throughout the paper.

Based on these characteristics, some networks exhibit small-world properties when compared to similar random networks (Milgram, 1967; Watts & Strogatz, 1998; Steyvers & Tenenbaum, 2005). Humphries and Gurney (2008) introduced a quantitative measure called smallworldness *(S)* aimed to assess the small-world properties of a network. This measure corresponds to the ratio of the clustering coefficient (CC) to the average shortest path length (ASPL). The network is considered smallworded when $S$ is higher than 3 (Humphries & Gurney, 2008). A small-world network when compared to a random network of equal size and number of edges would have a lower average shortest path length but a higher clustering coefficient (Erdös & Rényi, 1960; Watts & Strogatz, 1998; Steyvers & Tenenbaum, 2005; De Deyne & Storms, 2008). Smallworld networks represent a vast collection of models that can be found in multiple real-world phenomena.

**Small-world structures and semantic networks**

Milgram (1967), in an experiment on social networks tested the principle according to which people are on average separated by only six social connections. He asked a group of participants (origin node) to send a letter to a stockbroker (destination node) using their own social connections. Results suggested a small degree of separation between any two people living in the United States at the same time. According to Milgram (1967), this suggested an organisation of social connections according to a small-world structure.

Other studies later confirmed theses small-world properties in social networks (e.g., Watts & Strogatz, 1998; Kitsak et al., 2010), epidemiology of infectious diseases (e.g., Keeling & Rohani, 2008), spread of computer viruses (e.g., Pastor-Satorras & Vespignani, 2001), protein interaction (e.g., Bork et al., 2004), gene expression (e.g., van Noort, Snel & Huynen, 2004), neural (e.g., Basset & Bullmore, 2017) and semantic networks (e.g., Steyvers & Tenenbaum, 2005).

The use of semantic networks for the study of language corpora has a long history as evidenced by the development of influential computational models such as LSA

(Latent Semantic Analysis, Landauer & Dumais, 1997). It models the assumption that words that co-occur in similar contexts share the same meaning (Firth, 1968). This corpus approach relies on the mining of large text corpora that provide a good representation of the linguistic environment from which language is acquired and processed (De Deyne et al. 2016). The use of co-occurrence to define meaning is limited however because it does not distinguish between similarity-based co-occurrences and purely verbal associates as it is the case for psycholinguistic models of semantics. The use of word associations collected from human participants can address this limitation by providing large-scale networks to study the mental lexicon from data that are psychologically plausible.

The word association task is a simple yet powerful tool that can generate a large amount of data to gain insights into the organization of the mental lexicon. Traditionally, these data are used to create word association norms that are applied in factorial designs to study varied cognitive phenomena such as visual word recognition, memory recall, and semantic and lexical decision tasks (Balota, Cortese, Sergent-Marshall, Spieler & Yap, 2004; Hutchison, 2003; McNamara, 2005). The modelling of these data using network science not only is interesting because it provides psychologically plausible data or the ability to explore semantic memory on a larger scale than what is possible to achieve with factorial designs but is also theoretically sound. This technique is embedded in the seminal work of Collins and Loftus (1975) who established the network structure of the mental lexicon. In addition to the structural properties, semantic networks have the ability to model functional properties such as the spreading activation process that takes place when a node is activated in the network and all nodes that are connected to it are activated in turn. It is also possible to consider the weights and paths length in a directed network to model activation decay as a function of time and distance (Collins & Loftus, 1975; Den Heyer, & Briand, 1986; Kenett, Levi, Anaki & Faust, 2017).

Many studies have investigated the above-mentioned characteristics of semantic networks (Motter, Moura, Lai & Dasgupta, 2002; Sigman & Cecchi 2002; Steyvers & Tenenbaum 2005; Bales & Johnson 2006; Borge-Holthoefer & Arenas 2010; Choudhury & Mukherjee 2009; Fukś & Krzemiński 2009; Veremyev, Semenov, Pasiliao & Boginski, 2019).

Steyvers and Tenenbaum (2005) analysed semantic networks based on behavioural word associations and large text corpora. They showed that they all have a small-world structure characterized by sparse connectivity, short average path lengths, and a strong clustering coefficient. These small-world properties reflect an organisation of the lexicon in highly connected hubs that allow efficient information distribution and contributes to the robustness of the structure as a whole (Borge-Holthoefer, Moreno, & Arenas, 2012; De Deyne et al. 2016). It also provides insights into the network growth mechanisms as newly acquired concepts join tightly connected hubs over time (Steyvers, and Tenenbaum, 2005). These results were later replicated in Dutch (De Deyne et al., 2008), German and Spanish (Borge-Holthoefer & Arenas, 2010), Hebrew (Kenett et al., 2017) and most recently in Persian (Karimkhani et al., 2021) showing that these properties still hold for Semitic languages.

**The present work**

In the present work, we applied techniques of semantic network analyses on behavioural abstract and concrete word associations previously collected from French native speakers (Lakhzoum et al., 2021). The first aim of this work is to replicate the network analyses and small-world properties found by Steyvers and Tenenbaum (2005) and generalise their results to the French language. A second aim is to explore potential differences in network properties when separating abstract and concrete word associations in two distinct networks. This is, to the best of our knowledge, the first attempt to study the characteristics of abstract and concrete networks and the first attempt to generalise small world properties to the French language.

## METHOD

### Material and procedures

A detailed description of the French word association data is provided in Lakhzoum et al., 2021. The word associations were based on 1100 cues with varying levels of concreteness from very abstract to very concrete on a 7-points scale. They were collected in an online experiment among 1200 French native speakers. This resulted in a dataset of 92000 responses. Following the procedure of De Deyne et al. (2019), only the responses suggested by at least two participants were kept. For the remaining data, a weighted adjacency matrix was computed. The cue words and associated responses were used to create the nodes in the networks. The links between cue and responses were used as the directed arcs and the association strength values were represented in the weights of the arcs. The whole network was later divided into the concrete and abstract networks based on the median level of concreteness of the cues.

## RESULTS

Semantic network analyses do not rely on techniques of hypothesis testing, as it is the case for factorial designs (Moreno & Neville, 2013). The insights about the data are drawn from comparisons of directed and undirected networks with random networks of comparable size, density and mean connectivity <k>. More specifically, the benchmark random networks used to estimate smallworldness (ASPL and CC parameters) were generated with the same size and mean connectivity that provided the parameters $ASPL_{random}$ and $CC_{random}$ (see Table 1). Most of the literature on semantic networks relies on unweighted, undirected networks. However, considering the present application in word association data, the direction of the edges and the frequency of the association are of importance. The following analyses are based on directed and undirected weighted graphs. The variable *n* refers to the number of nodes corresponding to the word association dataset (Lakhzoum et al., 2021). The whole network corresponds to the graph representation of the entire dataset (see Figure 1 for a representation of the whole network; see Figure 2 for a more focused visualisation of the network)

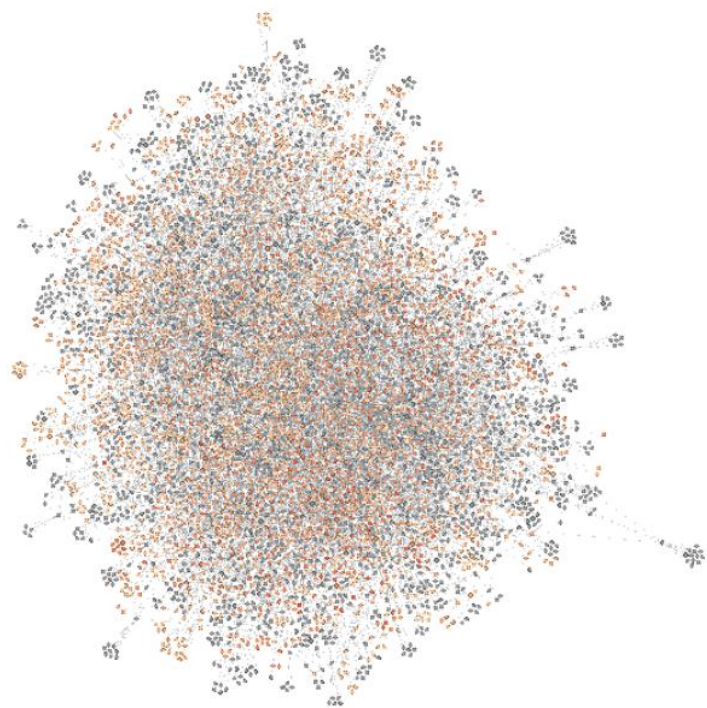

**Figure 1.** Visual representation of the whole network. Each point represents a word. The colour gradient from grey to red represents the distance between connected nodes the greyer the farther and the redder the closer.

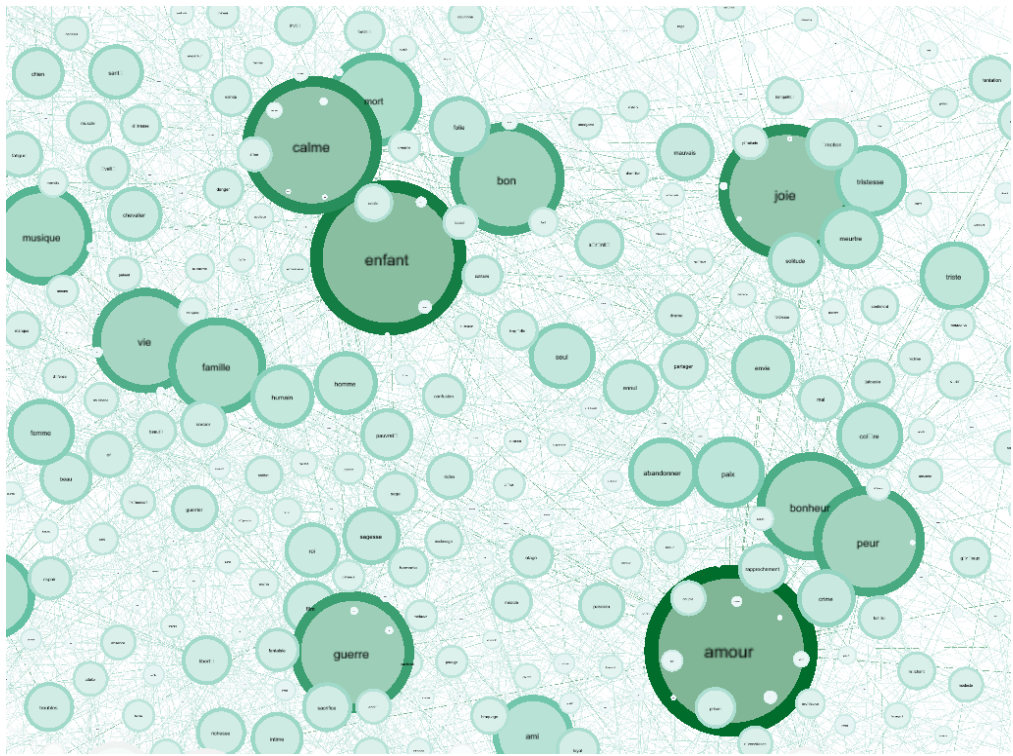

**Figure 2.** Visualisation of a portion of the network with a representation of the nodes degree were the larger and the darker green a node is and the more connection it has. The node with the most number of connection on this portion is the French concept "amour" or *love* in English. One of the nodes with the smallest degree is the French concept "ennui" or *boredom* in English.

Based on our previous study, we were able to determine that concrete and abstract cue words tend to elicit responses of the same level of concreteness (Lakhzoum et al., 2021) which justified our choice to explore both levels in separate networks. The concrete and abstract networks correspond to the datasets for the concrete and abstract cue words respectively.

The first parameter of interest is the diameter (D) of the network that defines the maximum distance between two nodes. It provides information about the macrostructure of the network and an estimation of the overall connectivity as it relates to the density parameter (d). For the whole network, the diameter shows that the most distant nodes are separated by 20 (directed) and 9 (undirected) connections. In the concrete network, the most distant nodes are separated by 24 (directed) and 9 (undirected) connections. Finally, in the abstract network, the most distant nodes are separated by 19 (directed) and 10 (undirected) connections. These diameters are overall higher than the ones found in the Leven dataset (De Deyne et al., 2008) which is explained by the higher density of their network.

**Table 2.** Network statistics for the directed and undirected associative networks for the whole, concrete and abstract datasets.

| variables | Whole network | | Concrete network | | Abstract network | |
| --- | --- | --- | --- | --- | --- | --- |
| | directed | undirected | directed | undirected | directed | undirected |
| n | 4,758 | | 2,961 | | 2,827 | |
| D | 20 | 9 | 24 | 9 | 19 | 10 |
| d | $5.10^{-3}$ | | $6.10^{-3}$ | | $7.10^{-3}$ | |
| ASPL | 7.07 | 4.90 | 8.46 | 5.13 | 7.20 | 5.00 |
| $ASPL_{random}$ | 10.00 | 7.05 | 10.21 | 6.63 | 10.13 | 6.58 |
| $k_{in/out}$ | 2.46 | - | 1.97 | - | 2.07 | - |
| <k> | 4.92 | 4.92 | 3.94 | 3.94 | 4.13 | 4.13 |
| CC | 0.15 | 0.07 | 0.15 | 0.07 | 0.14 | 0.06 |
| $CC_{random}$ | 0.0005 | 0.0005 | 0.001 | 0.0008 | 0.0007 | 0.001 |
| S | 682 | | 925 | | 657 | |
| Q | 0.62 | | 0.70 | | 0.64 | |

Note. n = the number of nodes; D = the diameter of the network; <k> = the average number of connections; ASPL = the average shortest path length; CC = clustering coefficient; $ASPL_{random}$ = the average shortest path length with random graph of same size and density; $CC_{random}$ = the clustering coefficient for a random graph of same size and density.

The density of the network serves two purposes. Firstly, it allows to create a comparable random network to provide a benchmark for small-world analyses. Secondly, it serves to check the sparseness of the network. The sparseness of the whole network is shown by a very low density of $5.10^{-3}$. This means that about 0.05% of all possible associations between words are represented. When comparing the density of the concrete and abstract networks, the concrete network presents an even lower density, which is in accordance with its higher diameter. This extreme sparseness can be explained by the variety of the dataset in terms of levels of concreteness of the cues. We expect that the concrete and abstract data share very few connections and belong to separate clusters. The sparseness of these networks -very low density- is the first mark of their small-world structure.

Based on the average shortest path length (ASPL), there was an average of 7.07 (directed) and 4.90 (undirected) steps from one node to another in the whole network. This average number of steps was higher in the concrete network (8.46 for the directed and 5.13 for the undirected) compared to the abstract network (7.20 for the directed and 5.00 for the undirected). This pattern is in accordance with the diameter. A series of random networks were generated with comparable size *n* and mean connectivity *<k>* to serve as benchmark for smallworldness estimation. All networks had the same density as their corresponding random benchmarks. According to Table 2, the ASPL statistics were consistently smaller for the actual networks compared to the randomly generated ones. A small ASPL is a fundamental characteristic of a small-world structure.

The clustering coefficient is a measure of the probability that associates of a word are also connected to each other. This parameter is evaluated against a comparable random network. Table 2 shows that associates of a word in the directed networks were also connected to each other about 15% of the times for the whole and concrete networks and 14% of the times for the abstract network. For the undirected networks, this probability dropped to approximately 6-7 % of the time. The clustering coefficients of the whole, concrete and abstract networks were very high compared to the ones found for the benchmark random networks that ranged from 0.05 to 0.1% probability to find a link between two nodes that have a neighbour in common. The clustering coefficient is among the most important indicator of a small-world structure. It suggests that when additional nodes are introduced in the network, they join a tightly connected hub, which insures the stability of the entire network.

Taken together, the sparseness, average short path lengths and clustering coefficient show that the present networks exhibit the same small-world properties found by Steyvers and Tenenbaum (2005). A final measure established by Humphries and Gurney (2008) was computed for each network and their corresponding random networks to evaluate their smallworldness. Based on this parameter, Humphries and Gurney (2008) indicated that an S measure > 3 indicates a small-world structure. All the networks presented a very high S measure.

Finally, the modularity measure (Q) introduced by Newman (2006) indicates the extent to which a network is composed of sub-communities. Communities are non-overlapping groups of nodes with a high number of within community connections and a low number of between-community connections. The concrete network is more modular (Q = 0.70) compared to the abstract network (Q = 0.64).

Network analysis is rooted in the seminal work of Collins and Loftus (1975) who described the spreading of activation according to which the activation of concepts in memory spreads and activates other closely related concepts. This process is fundamental in cognitive science for its ability to explain many phenomena related to the mental lexicon (see Siew et al., 2019 for a review). Given the structural differences we have found in the concrete and abstract networks, it is fair to assume that the two types of concepts would exhibit different patterns of spreading scores. In the same way that spreading activation mimics the diffusion of discrete events such as the spread of disease among individuals, it is possible to identify initial "infected" nodes and their ability to spread the discrete event which is modelled (Siew et al., 2019). Salavaty, Ramialison and Currie (2020) have proposed a computational method to identify the most influential nodes in a network that have a higher spreading scores for each node. Spreading scores are an indication of the capacity for a node to spread information within the network. The algorithm was used for each abstract and concrete network and generated a vector of spreading scores for each node. Comparison of spreading scores for abstract and concrete nodes respectively showed that abstract nodes (M=2.56; SD=4.75) have a lower spreading score compared to concrete nodes (M=3.72; SD = 6.57) – $t(5390) = 7.70$; $p < 0.0001$. These results suggest that the abstract nodes in the network are less influential and have a lesser ability to spread information within the network compared to concrete nodes.

## DISCUSSION

In the present work, the data obtained from word associations was represented as several networks using graph-theory based principles. This aimed at extracting structural and organisational patterns of the mental lexicon that would not be possible to gain from the raw dataset (De Deyne & Storms, 2008; Csardi & Nepusz, 2006; Kolaczyk & Csardi, 2014). The whole network represented the entire dataset. Next, based on a previously established pattern of cues eliciting responses of the same level of concreteness (Lakhzoum et al., 2021), the whole network was divided into a concrete and an abstract network by using the median concreteness level to separate them. A first series of analyses used random networks of a similar density and connectivity as benchmark (Lerner, Ogrocki, & Thomas, 2009). Comparisons between the present networks and comparable random ones showed that the networks exhibited a small ASPL and a high clustering coefficient characteristic of a small-world structure (Watts & Strogatz, 1998; Steyvers & Tenenbaum, 2005; De Deyne & Storms, 2008). The small world structure provides valuable information not only about the structural characteristics of the networks but also about the way the networks will expand with the addition of new nodes. In a small-world network, when new nodes are included they join an already established cluster of existing nodes. This means that the network does not expand proportionally with the addition of new nodes. This small-world effect insures that networks present a highly robust architecture that is resistant to change and disturbances (Peng, Tan, Wu & Holme, 2016).

When comparing the topological properties of the concrete and abstract networks, the concrete networks seems more spread out (higher ASPL), more connected (higher S measure), more modular (higher Q measure) than the abstract network. This pattern of densely connected but loosely interconnected nodes for the concrete network concurs with the pattern found in association strength with concrete cues eliciting stronger associates compared to the abstract ones. It also concurs with the dichotomy established by Paivio (Paivio, Yuille & Madigan, 1968; but see Della Rosa, Catricalà, Vigliocco, Cappa, 2010) between concrete and abstract concepts with concrete concepts being more highly connected to a smaller number of contexts while abstract concepts are more loosely connected to a higher number of contexts. This pattern of abstract concepts was confirmed in previous studies on feature-listing norms (Recchia & Jones, 2012) and now on a larger scale as evidence by a smaller modularity measure in the abstract network.

Taken together with previous findings, these local node statistics suggest that abstract and concrete concepts do not differ in their pattern of connectivity but rather on the strength of these connections with concrete concepts presenting stronger association compared to abstract concepts (Lakhzoum et al., 2021). This pattern of weaker connection for abstract nodes compared to concrete nodes is further reflected in their respective spreading scores with abstract nodes presenting a lesser ability to spread information in the network compared to concrete nodes. These scores are reflective of the position of the nodes in the network and influenced by clustering and modularity as well as the weights of their connections. This suggests that abstract concepts, due to their more diffuse pattern of organisation and weaker connections given by associative strength from the word association data have a lesser ability to spread semantic activation in the mental lexicon. This is in accordance with previous findings according to which abstract concepts present a processing disadvantage compared to concrete concepts as evidenced by the longer response times they elicit (see Borghi, Binkofski, Castelfranchi, Cimatti, Scorolli & Tummolini, 2017).

Taken together, these characteristics suggest that the local organisation of abstract and concrete concepts is similar while their structural position within the network shows a different pattern with concrete concepts presenting a more modular pattern compared to abstract concepts.

FUTURE WORK

The present work represents a first attempt to study the structural dichotomy in the organisation of concrete and abstract concepts in the mental lexicon. A future project will have to replicate the same analyses using an English word association dataset such as the SWOW-EN database of De Deyne and colleagues (2019). This will allow for a replication on a higher scale because the database contains 12.000 cues for more than 4 million responses. Realising the same network in English will also provide a baseline comparison for the research community and the opportunity to compare between English and other languages. By cross-referencing with concreteness ratings norms (Brysbaert, Warriner & Kuperman, 2014), it will be possible to understand the same structural organisation of abstract and concrete concepts on a larger scale.

Additionally, it must be kept in mind that the present network representation is not a neural network and does not yield any insights into the mapping of abstract and concrete concepts representation in the human brain. Network science can be applied to study brain connectivity based on the same techniques and parameters used in the present work (Siew et al., 2019 for a review). Based on recording of brain activity during a word association task for abstract and concrete concepts, it will be possible to map the cognitive patterns obtained in the present work to the neural substrates that generated said cognitive patterns.